\documentstyle[aps,epsf,prl,twocolumn]{revtex}
\begin{document}
\twocolumn[{
\draft
\title{Structure, Stresses and Local Dynamics 
in  Glasses}
\author { Tamar Kustanovich and Zeev Olami}
\address{Department of~~Chemical Physics,\\
 The Weizmann Institute of Science,
Rehovot 76100, Israel}
\maketitle
\widetext
\begin{abstract}
\leftskip 54.8pt

\rightskip 54.8pt
The interrelations between short range structural and elastic aspects in glasses
 and glass forming liquids pose important and yet unresolved questions.
In this paper these relations are analyzed for mono-atomic glasses and stressed liquids with a short range repulsive-attractive pair potentials.
Strong variations of the local pressure are found even in a zero temperature glass, whereas
the largest values of pressure are the same in both glasses and liquids.
The coordination number $z(J)$ and the effective first peak radius depend on the local pressures $J$'s.
A linear relation was found between components of site stress tensor
and the local elastic constants. A linear relation was 
also found between the trace of the squares of the local frequencies and the local pressures. Those relations hold  for  glasses at zero temperature and for liquids.
We explain this by a relation  between the structure and the potential terms.
A structural similarity between liquids and solids is manifested by similar dependencies of the coordination number on the pressures.

\end{abstract}

\leftskip 54.8pt
}]
\narrowtext
%
\section{Introduction}
Short range order in glasses and liquids has a set of different interrelated aspects: structure, local stresses and the elastic properties (e.g. local frequencies). Though a lot of attention has been devoted to the structural point
of view and to frustration effects [1-7],
 almost no attention was given to the local
elastic properties \cite{SC91} and to the existence of a wide range
 of local stresses in such systems
\cite{egami,Alexander,Kust}. In a liquid state,
there is a great variety in the local pressures felt by atoms, as
a result of the variations in local environments\cite{egami}.
When a glass is created from the liquid state, its atoms will reach one of 
the local minima of energy (defined by zero force), but not the global energy minima (corresponding to the crystal). 
Therefore, the local pressures in the 
compressed liquid will not be fully relaxed in the quench. 
Furthermore, the existence of different stressed local environments 
will result in varying local elastic constants 
in compressed liquids and glasses.

Crystals and glasses are notably similar, since both have a fixed structure
around which the  zero point  dynamics can be discussed.
Yet, there is a basic structural difference between them. 
In crystals, due to the crystalline order, there is a unique lattice with
 well defined distances between sites
and periodic long range order that corresponds to the structure at minimum energy.
The dynamical properties and elastic constants are obtained by expanding the energy around that unique
 reference lattice.   
In glasses there is only short range order, which is determined by
the zero force condition and by the fluctuations in the local pressures. On larger scales, the structure is  homogeneous.

It is important to realize that the  knowledge of structural aspects of the glass 
cannot supply any information about the glassy dynamics. 
Furthermore, to understand the glassy structure one has to 
consider information related to the internal stresses.
Hard sphere glasses can supply a useful first order description of structure in glasses\cite{zallen}.
However, packings with very similar structure
 can have a very different elastic description, so knowledge of the forces is essential.  Even if some glassy structure is known, 
a redefinition of the nearest neighbors interactions 
  can completely modify the spectra.
In this paper we explore this problem, focusing on the relations between
the local dynamics, stresses and structure.
We have found that the fluctuations in structure and 
in the force are combined together to create a linear relation between
 the local elastic constants and the local pressures.

This kind of analysis can enhance the understanding of
structure in glasses.  It can be used to predict the variability of stress
and elastic constants in glasses, which can then be used to give predictions 
on the global spectra and other glass characteristics. 

In this paper we discuss short range mono-atomic glasses (SRMA glasses) and stressed liquids.
In section III it is shown that local pressures are distributed in
 a wide stable distribution, which
is related to the pair distribution function.
There is a common upper cutoff
for the local pressure distributions in glasses and liquids.
We have found that the variations in pressure and structure are correlated
and  that 
features of the non-stable liquid state are sustained in the glass.

Section IV describes correlations between stresses and local dynamics.  
We have found a linear relation between local pressures and
the elastic constants. This relation can be explained by combining
the structural information discussed in section III with  knowledge of the 
potential. Thus, this linearity is a strong constraint on the
structure of 6-12 Lennard-Jones glass. Furthermore, it seems to be a general feature of SRMA glasses (see section V).
%
%
\section{Elastic Description}
First, we define our basic points of interest.    
Consider a solid at equilibrium state at zero temperature. 
There are no forces on atoms and no external stresses on the boundaries. 
In such case, the average stress and, in particular, the average pressure are zero.
Note that to ensure absence of stresses on boundaries, one needs to consider only potentials with attracting and repulsive terms (e.g. 6-12 Lennard-Jones potential) with density adjusted so that total pressure equals zero. In crystals, the average stress (pressure) and the deviations from it are zero. {\bf This is not so in amorphous solids}: 
even if the amorphous system under consideration is left with a zero average stress, the local stresses can be very different from zero \cite{Alexander,egami,Kust}. 

Consider a discrete system where atoms $i$ and $j$ interact through a 
pair potential $U(R_{ij})$, where 
$R_{ij}$ is the distance between two atoms.
The forces between particles are defined by:
\begin{equation}
\vec{f}_i = \sum_{i} U'(R_{ij})\frac{\vec{R}_{ij}}{|R_{ij}|}
\end{equation} 
where $U'(R_{ij})=d U_{ij}/dR_{ij} $ and $\vec{R}_{ij}=(X_{ij}, Y_{ij}, Z_{ij})$. In a glass at zero temperature:
\begin{equation}
\vec{f}_i= 0
\end{equation}

The site stress tensor can be defined as \cite{Alexander,egami}
\begin{equation}\label{S}
{S}_{i}=\sum_{j} S_{ij}= -  \sum_{j} \frac{U^{'}_{ij}}{R_{ij}} \left(\matrix{X^{2}_{ij} &  X_{ij} \: Y_{ij}& X_{ij} \: Z_{ij}\cr
X_{ij} \: Y_{ij} & Y^{2}_{ij}& Y_{ij} \: Z_{ij}\cr
Z_{ij} \: X_{ij} & Z_{ij} \: Y_{ij}& Z^{2}_{ij}\cr }\right)
\end{equation}

The average stress field around atom $i$ can be defined as the
 average stress in a cell of volume $V_{i}$ 
\begin{equation}
\sigma_i = \frac{S_i}{ V_i} 
\end {equation}
Usually this cell is supposed to be the Voronoi cell, however this is not a 
necessary assumption  \cite{Alexander,egami}. At equilibrium, it follows that   
\begin{equation}\label{sig}
<\sigma>=\sum_{ {all-atoms}} \sigma_i \equiv 0
\end{equation}
However there is no such constraint on the various $\sigma_i$. 
The same requirement holds for the pressures.
 We define  the term 
\begin{equation}\label{J}
J_i =  \sum _j U'(R_{ij})
\end{equation}
as the local pressure.
The distribution function of the local pressures is denoted as
$P(J)$.

The local dynamics of an atom is defined by the 
 matrix $A^i=\sum A^{ij}_{\alpha \beta}$ 
which describes the second derivatives of 
energy in particle $i$
while other atomic positions are fixed. The 
$A^{ij}$'s are defined as:
$A^{ij}_{\: \: ,\alpha \beta}=\partial_{\alpha}\partial_{\beta} U^{ij} $,
where $\alpha$ and $\beta$ denote components of a vector. 
For radial pair potential
\begin{equation}\label{A}
{A}_{i}=  \sum_{j} \left[ \frac{U^{''}_{ij}}{R_{ij}^{2}} - \frac{U^{'}_{ij}}{R_{ij}^{3}} \right]
\left(\matrix{X^{2}_{ij} &  X_{ij} \: Y_{ij}& X_{ij} \: Z_{ij}\cr
X_{ij} \: Y_{ij} & Y^{2}_{ij}& Y_{ij} \: Z_{ij}\cr
Z_{ij} \: X_{ij} & Z_{ij} \: Y_{ij}& Z^{2}_{ij}\cr }\right) + \frac{U^{'}_{ij}}{R_{ij}} \mbox{I}
\end{equation}
where $U^{''}_{ij}=d^2 U_{ij}/dR_{ij}^{2}$ and I is identity matrix. Note, that there are terms in A which are related to the first gradients \cite{Alexander}.
The trace of this tensor is:
\begin{equation}\label{K}
K_i = trace(A_{i}) =  \sum_{j} [ U^{''}_{ij}+ \frac{2 U^{'}_{ij}}{R_{ij}}]
\end{equation} 

A liquid of the same volume has velocities and the zero force condition does not hold. 
 However it is still interesting to use  the previous elastic descriptions. 
The stress, pressure and local elastic
frequencies can still be measured, even though they do not have such a direct relationship 
to the spectra.  Doing so, we ignore the effect of velocities on the pressure.

%
%
\section{Stress and Structure}
\subsection{Systems Studied}
We considered  systems with 4096 atoms interacting though 6-12 Lennard-Jones  potential, using the values $\varepsilon/k_b=120$ and $\sigma=0.34 nm$, which correspond to Argon. Periodic boundary conditions were assumed.
 The density $\rho=1.763 gr/cm^{3}$ was chosen to ensure no external pressures at equilibrium.
 The samples were prepared from a compressed liquid at 120K, by using
a molecular dynamics simulation \cite{Com}, either by a steepest decent quench to zero
 temperature glass, or by coupling to a heat bath at temperatures of 1 and 8K. 
Details of glass preparation by various methods are summarized elsewhere \cite{Kust}. 
All those glasses were quenched to zero temperature, to ensure the existence of an elastic description. 

We calculated the local $S_{i}$, $J_{i}$, $A_{i}$ and $K_i$ (eq. 3, 6, 7 and 8) on various samples. Some initial calculations were made with other potentials as well \cite{remark-1} (see comments in section V). 

\subsection{Local Stresses and Structure}
A fully relaxed glass with zero global pressure has large variation of local pressures $J$'s. 
Therefore, we first look at the width of distribution and discuss the
essential features of the local pressures in our system.
The distributions of $J$ for a glass at T=0
 and for liquid at $T= 120K$ are presented in Figure 1. 
\begin{figure}
\epsfxsize=6truecm  
\hspace*{0.4truecm} \epsfbox{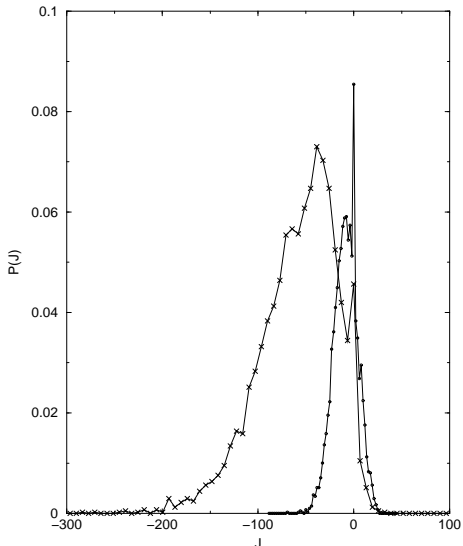}
\caption{$P(J)$ vs. $J$ for liquid at 120K (x) and for glass at 0K ($\cdot$). $J$'s  units are $\varepsilon/\sigma$, where $\varepsilon$ and $\sigma$ are Lennard-Jones parameters.}
\end{figure}

The distribution in glass, 
$P_{gl}(J)$, is wide, in the sense that its width
is of the same scale as the typical $J$'s for which the system will fracture
 (at uniform expansion)\cite{Kust}.
The average of $J$ is not zero even though the system 
is under a zero pressure (see eq. \ref{sig}). 
There is a sharp peak in $P_{gl}(J)$  at $J=0$.
This peak is due to about $1-2\%$ of particles that were found to be in an 
'ideal' equilibrium state (they are located in exact potential minima). 
Those minima are 
due also to neighbors which are far from the first peak. 
The preparation methods do not affect significantly the width of those distributions. In fact $P(J)$ is as stable as the radial pair distribution function $g(r)$.
Changes occur only due to crystallization. 
  
The liquid distributions $P_{liq}(J)$ 
 are wider \cite {egami} and the zero frequency peak is much smaller.
 Yet they
are similar in form to the glassy distributions.  
A surprising aspect is that the highest local  pressure $J_{max}$ is 
{\bf identical} in the liquids and the glasses distributions.
It is impossible to generate a site with a  higher local pressure. 
We discuss this point later.
The shapes of $P_{gl}(J)$ and $P_{liq}(J)$ can be re-scaled on each other,
 using the relative width of the distributions (or the average pressure)
 and the maximal value $J_{max}$.

In glasses and liquids, the
 usual description of the radial structure is given through
the radial pair distribution function $g(r)$, which is defined by
 the average number of neighbors at distance $r$.
 This function is not sensitive to variations in the local quantities.
An integration of the type
$J_{av}=\int g(r) U'(r)dr$ will give only 
the average local pressure in the glass, but will not provide any
information about the variability of $J$ or of the other parameters.  
Conditional radial pair distribution  $g_J(r)$ is defined as
 the average number of neighbors at distance $r$ for a specific given $J$. 
Since 
$g(r) =\int P(J) g_J(r)dJ$, all the information about the variability of $J$ 
is given indirectly in $g(r)$.
However, the structural effects of the local pressures can be observed directly, using the conditional radial distributions $g_J(r)$.
\begin{figure}
\epsfxsize=6truecm 
\hspace*{0.7truecm} \epsfbox{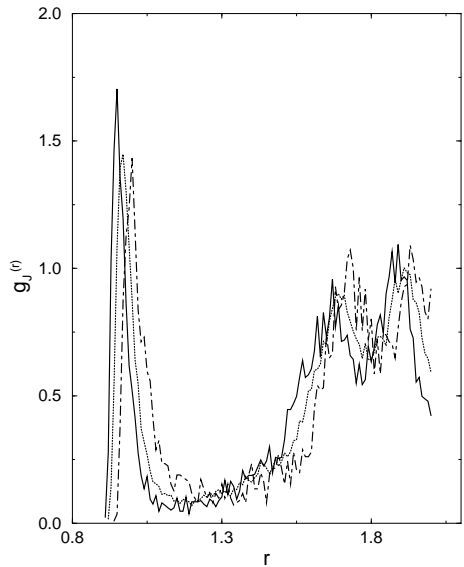}
\caption{$g_J(r)$ vs. $r$ for 3 averaged values of J - solid curve for compressed local environments,  dotted curve for low pressures (small values of $|J|$) and dashed-dotted line for stretched local environment (high values of J). $r$'s in units of $ 2^{1/6}\sigma$. }
\end{figure}

The conditional distribution functions $g_J(r)$
are shown in Figure 2.
As expected, $J$  is dominated 
by contributions from the first peak while there is an almost constant contribution
from further neighbors.
This is due to the fact that the pressures are much smaller at large distances and
the distribution is smoother. A reasonable approximation to $J$ is:
\begin{equation}
J = J_{fp}+ J_{bg}
\end{equation}
where we denote $_{fp}$ and $_{bg}$ as the first peak and further neighbors contributions,
 respectively.
 A numerical estimate for 
$J_{bg}$ is  $(11.5 \pm 1.5)\varepsilon/\sigma$.
Though the main effect on the local pressure
arises from the first peak in $g_J(r)$   
there are large structural effects in the second peak. 
This finding is related to midrange order in 
glasses, which was discussed extensively in the last decade 
\cite{eli,cnf98} and will be addressed  in a forthcoming
paper \cite{Olam99}.

The first peak is described by two major structural parameters: the coordination number and the average peak position, given by 
 the following definitions:
\begin{equation}
z(J) =\int_{fp} g_J(r) dr 
\end{equation}
\begin{equation}
\label{RJ}
R(J)_{av}= \frac{\int_{fp} g_J(r)r dr}{z(J)}
\end{equation}
where the cutoff of those integrals is defined by the end of the first peak.
Two major effects arise from the analysis of the dependence of $z(J)$ and
$R(J)_{av}$ on $J$: a shift of the average position of the peak $R(J)_{av}$  and a change in the coordination number $z(J)$. 
z(J) is presented in Figure 3. We have found a square dependence of $z(J)$ on $J$ for medium and high local pressures.
\begin{center}
\begin{figure}
\bbox {(a)\epsfxsize=3.4truecm\epsfbox{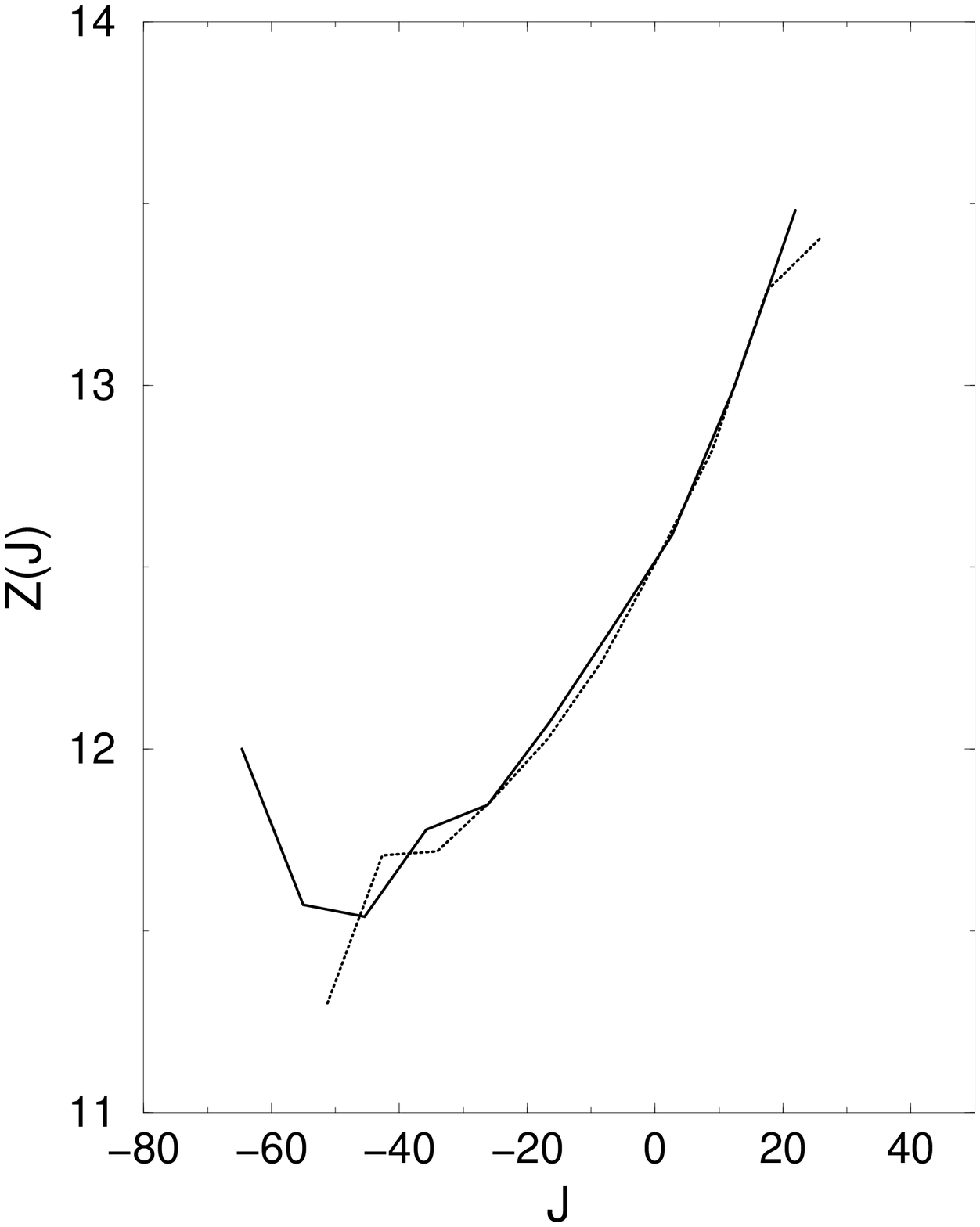}
(b)\epsfxsize=3.4truecm\epsfbox{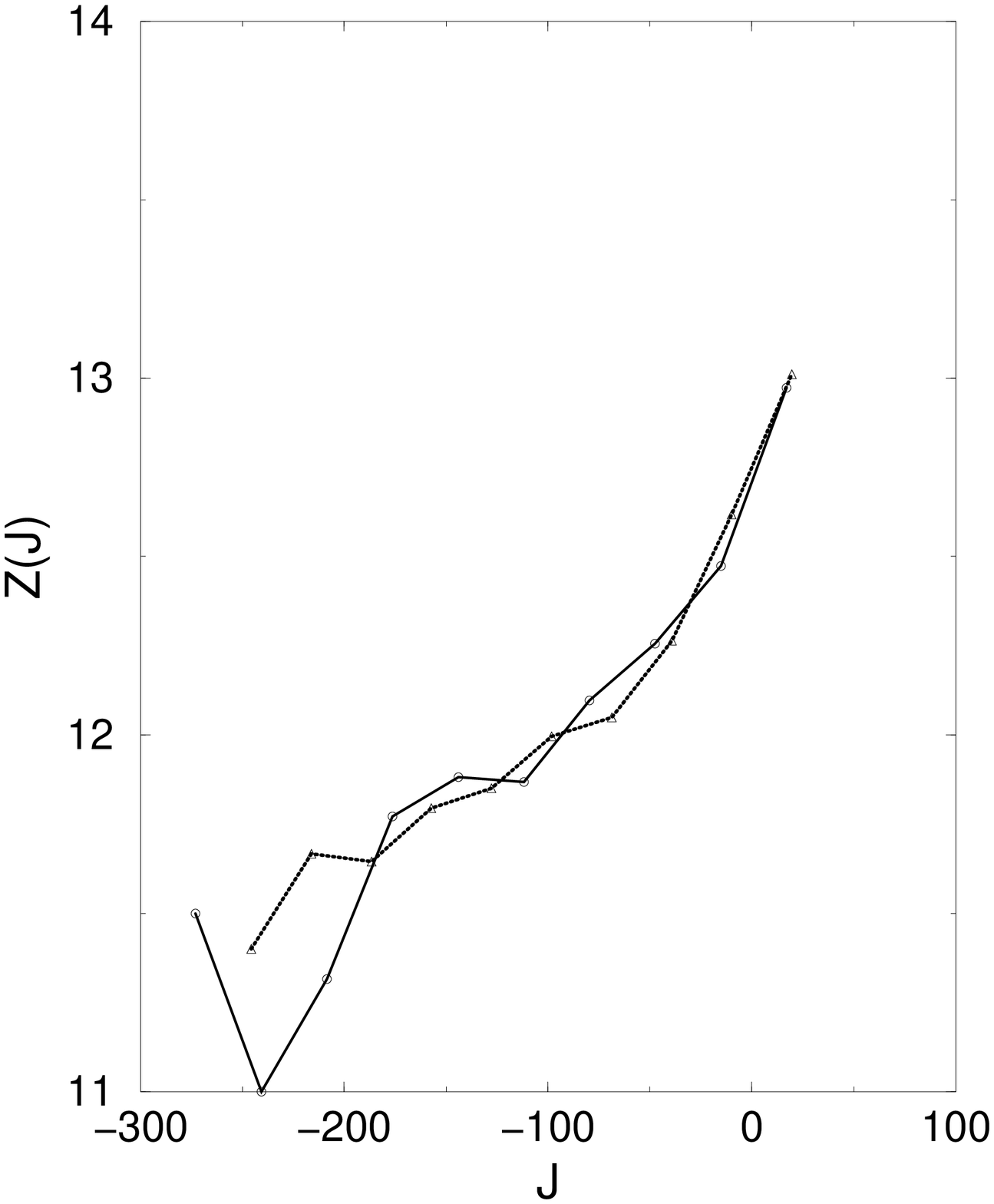}}
\vskip 0.2cm
\caption{
(a) $z(J)$ vs. $J$ for two different samples of glass at 0K. The plot has a good quadratic fit for $J>25$ (except for the upper point which is statistically meaningless as it represents $\sim 0.006 \%$ of atoms). (b) $z(J)$ vs. $J$ for two  samples of liquid at 120K, from which glasses at (a) where created.
}
\end{figure}
\end{center}
A  structural similarity between the glasses and liquids is observed
at high local pressures. 
For lower pressures the environments become denser and one observes stronger
fluctuations and lower coordination numbers.
The dependence of the average radius on $J$ is shown
 in Figure 4. 
As a somewhat better estimate for the  radius one can use $R(J)$,
defined as a solution of the self-consistent equation:
\begin{equation}\label{RJI}
J = z(J) U'(R(J))+J_{bg}
\end{equation}
This radius is also given in Figure 4.
Since $U'(R)$ increases  with distance, $R(J)_{av}$ and $R(J)$ grow with $J$ . 
\begin{center}
\begin{figure}
\epsfxsize=4.5truecm 
\hspace*{0.4truecm} \epsfbox{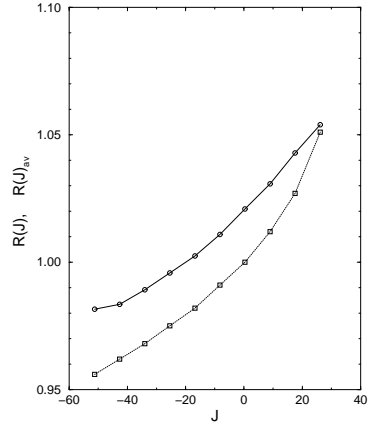}
\caption{Solid line with circles $R(J)_{av}$ vs. $J$ (eq. 11).
 Dashed line with squares 
 $R(J)$ vs. $J$ where  $R(J)$  is calculated from eq. 12.
}
\end{figure}
\end{center}
 The increment in $z(J)$
is not an obvious effect at all. One could argue for such effects from considerations of packing
 in the larger nearest neighbor coordination shell, but this will not explain the observed results. 
In fact we have obtained converse effects when other potentials were considered \cite{remark}, so it seems that there are various effects due to the potential shape.   
Notably the coordination numbers are all close to $12$, which is the 
so called  'icosahedral' packing [4-6]. However there are very strong variations. 
One can observe the same effects in the liquid state, where the rescaled $z(J)$
 still has a square dependence on $J$.
We note that the number of neighbors increases with
$J$ until it reaches a value of 13.5 and then one observes no more environments.
 The system cannot sustain larger $z(J)$ and $R(J)_{av}$'s 
because such an environment will collapse.

To understand this,
 consider the following one dimensional example. Suppose that
 an atom is attached to two opposite 
walls with a potential $U(R)$. If the walls are close enough there is
 only one stable solution with a negative
pressure. If the walls are far away there will be two stable solutions  near
  the walls and 
an unstable solution at the center. The criterion for the loss of stability of 
the central solution is $U''(R_{co})=0$, where $R_{co}$ is the distance to the wall.
 For the LJ potential $ R_{co} = \sigma (26/7)^{1/6}$, whereas the distance
 of zero force for the single pair interaction is $R_0 = \sigma 2^{1/6}$.
Once a particle's distance exceeds $R_{co}$, it will simply attach itself to another atom. 
A glance at the distribution function $P(J)$ shows that this scale is the exact cutoff 
for the distribution of the largest $J$'s. This mechanism is probably
the main  mechanism for two-level system dynamics for such glasses.
%
%
\section{Stresses and elastic constants}
Although the existence of correlations between the stress and
the local elastic constants is expected \cite{Alexander}, there is no prediction about their form.
We studied the dependence between tensors $A_i$ and $S_i$ and the scalars $K_i$ 
and $J_i$ (calculated from eqs. \ref{A}, \ref{S}, 
\ref{K} and \ref{J}, respectively). 
We have found linear relations between  $A_{i}^{\alpha \beta}$ and $S_{i}^{\alpha \beta}$:
\begin{equation}
 A_{i}^{\alpha \beta} =  \left\{ \begin{array} {rcc}
     a_1 S_{i}^{\alpha \beta} + b_1 &  &\alpha = \beta \\ 
     a_2 S_{i}^{\alpha \beta} + b_2 &  &\alpha \neq \beta \end{array} \right.
\end{equation} 
though no such relations are found between the specific 
$U'$ and $U''$ in eqs. \ref{A} and \ref{S}.
  Those results hold despite {\bf strong variability} between the different 
elements in $A$ and $S$.  A linear relation exists also between $K$ and $J$:
\begin{equation}
\label{KI}
 K_i =a J_i + b
\end{equation}
Those relations are found  both for glasses and for liquids.

The small variations in the middle curve for K (see Figure 5) indicate that 
eq. \ref{KI} is not an 'exact' mathematical identity.
Numerical estimates from eq. \ref{KI}  for one of the samples are $a=-0.24$ and $b=12.2$. When only non stressed terms in eq.\ref{A} are considered, $a=-0.21$ and $b=12.0$. The formal errors in the coefficients are $<1\%$. The numerical estimates from eq. \ref{KI} are $a=-0.235$ and $b=12.6$. Very close numerical estimates were obtained for other samples that were studied (deviations $<1\%$).
Results for specific eigenvalues are much more variable,
as can be seen in Figure 5.
\begin{figure}
\bbox {(a)\epsfxsize=3.4truecm\epsfbox{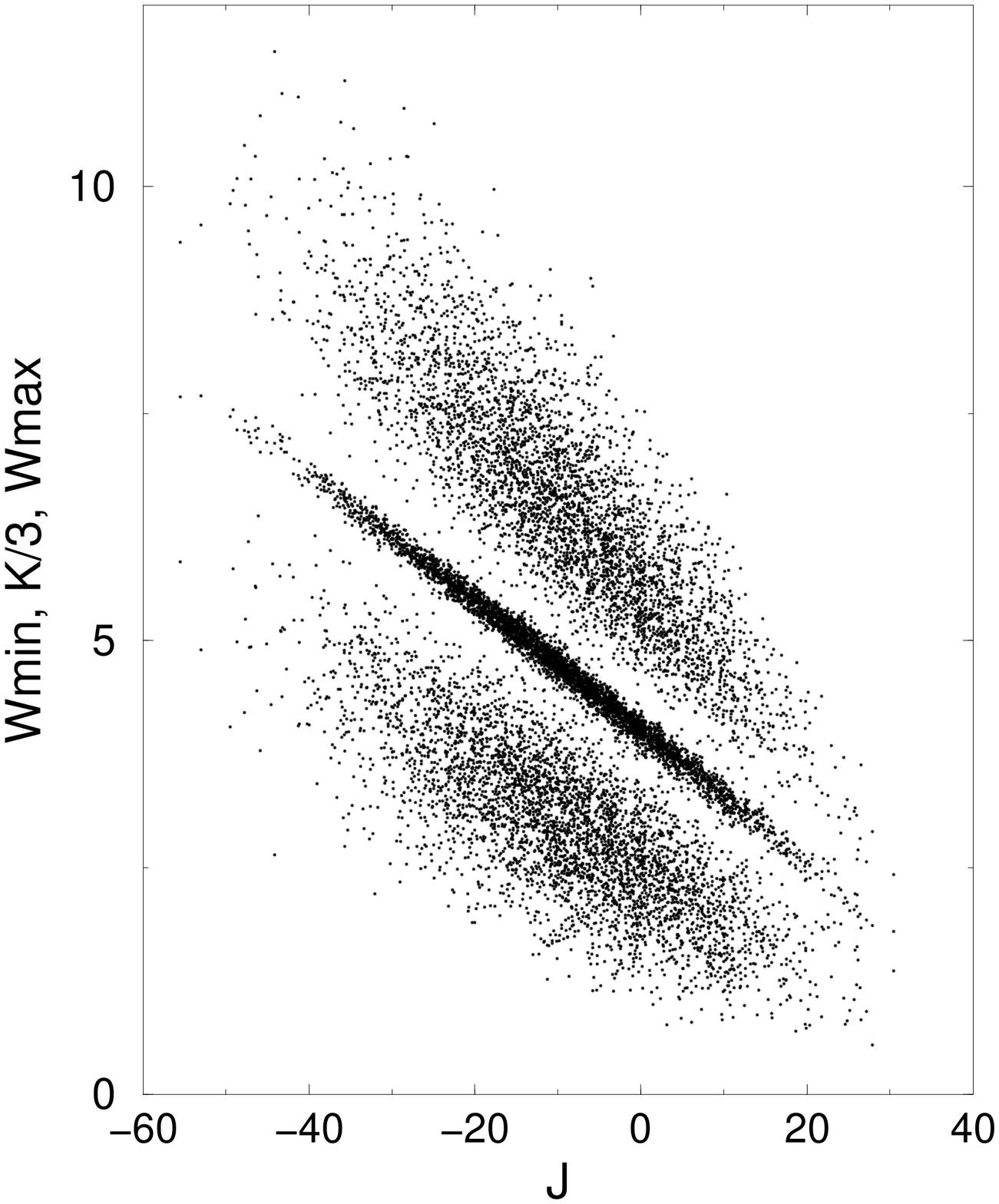}
(b)\epsfxsize=3.4truecm\epsfbox{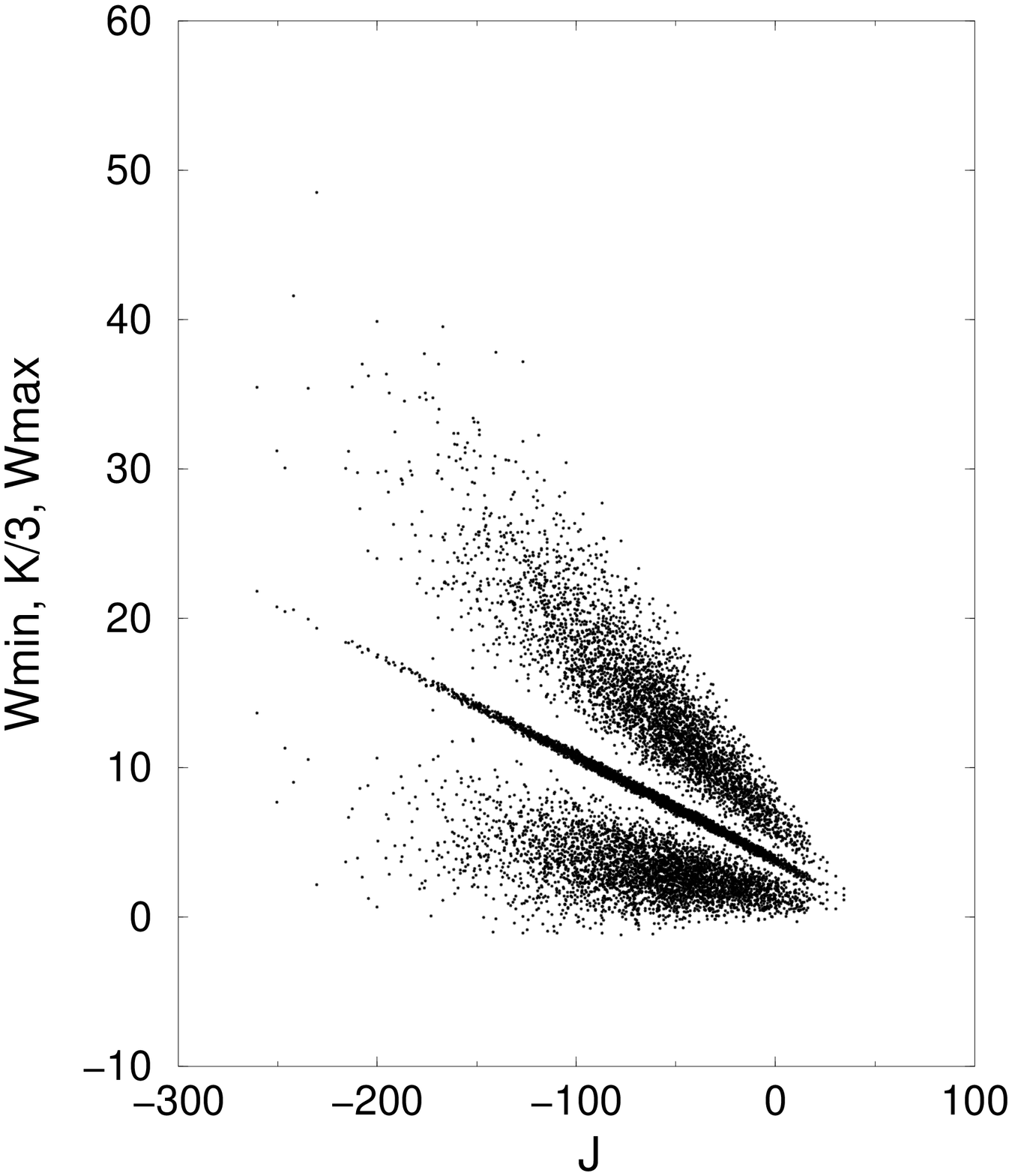}}
\vskip 0.2cm
\caption{
(a) Upper, middle and lower plots are, respectively, the largest ($W\mbox{max}_i$), the average
($K_i/3$) and the smallest ($W\mbox{min}_i$) eigenvalues of $A_i$ vs. $J_i$ for glass at 0K.  (b) $W\mbox{max}_i$, $K_i/3$ and $W\mbox{min}_i$ vs. $J_i$ for liquid at 120K. 
}
\end{figure}
The difference between liquids and glass is manifested by two  main effects: 
the width of the plots in Figure 5, and the variation of the definite eigenvalues of 
$A$. In the liquid state, fluctuations in values of the eigenvalues grow, so that 
many of the sites are unstable.  

As far as the distribution of local force constants is concerned,
the $K-J$ relation (eq. 14) provides the essence of the relations found in eq. (13). 
Therefore in the following we consider only the scalars $K_i$ and $J_i$. 
Linear fits for the $K$ vs. $J$ are shown in Figure 6, for the same glasses and liquids.   
\begin{figure}
\bbox {(a)\epsfxsize=3.4truecm\epsfbox{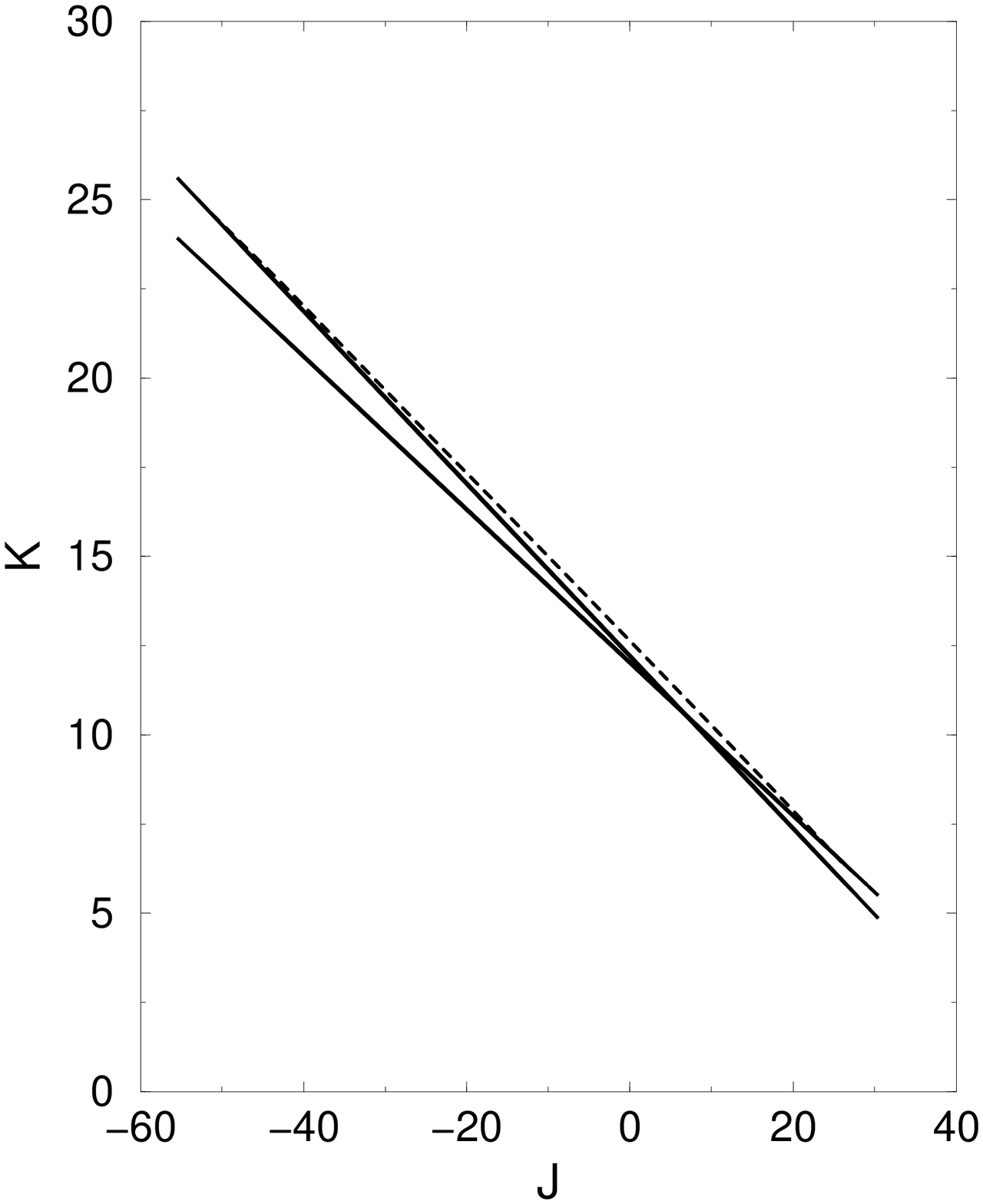}
(b)\epsfxsize=3.4truecm\epsfbox{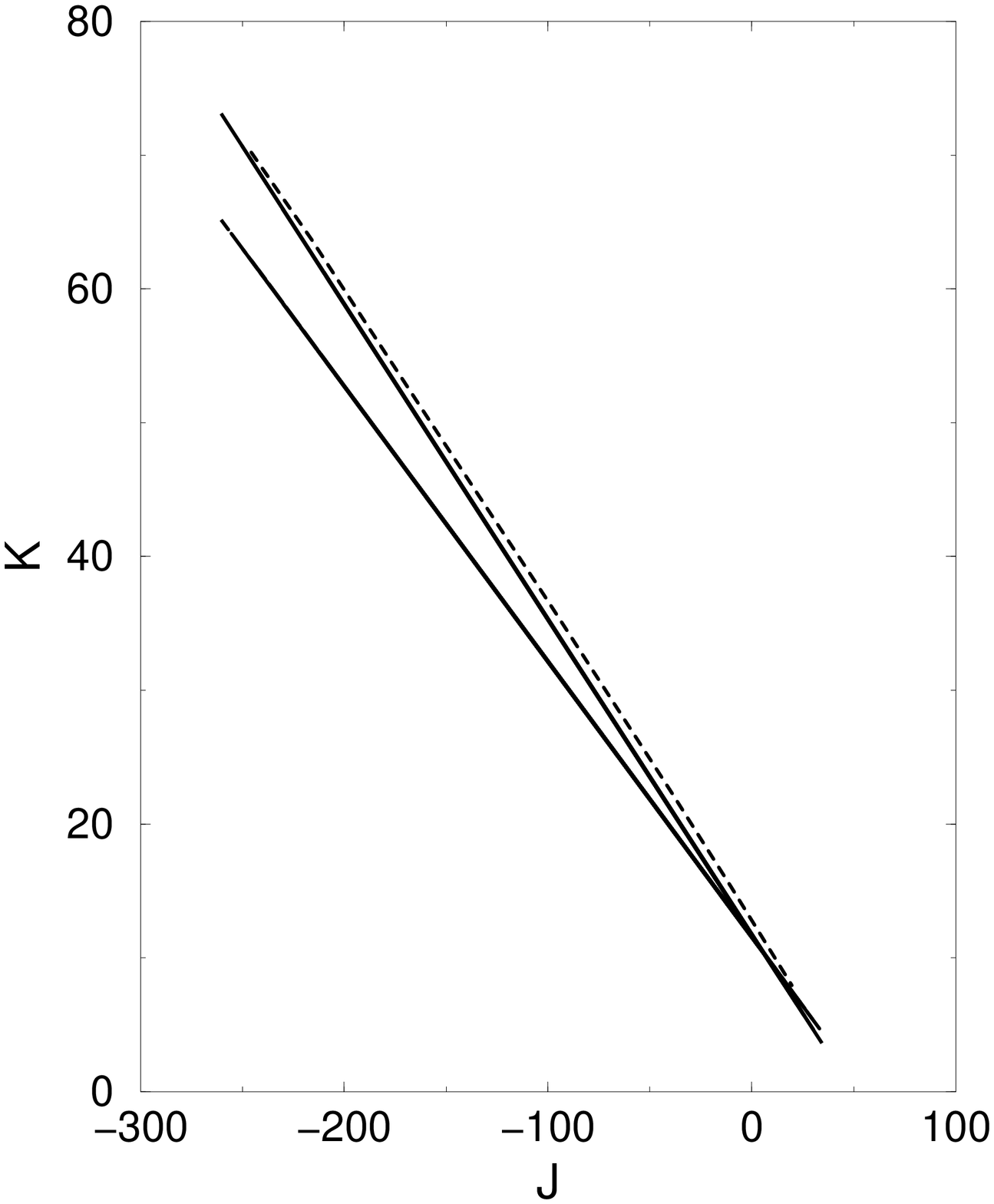}}
\vskip 0.2cm
\caption{
(a) $K$ vs. $J$ for glass at 0K. Solid line - linear regression of the 
$K_i-J_i$ curve calculated using the points in Figure 5. Long dashed line - linear regression of the $K_i-J_i$ calculated in the same way  without the stress terms in eq. 8. The width of the $K_i$-$J_i$ plots is very narrow, with formal error $<1\%$. Dotted line - K-J plot calculated form eq. 17. (b) $K$ vs. $J$ for liquid at 120K. Solid, long dashed and dotted lines calculated as on (a).
}
\end{figure}

How can one explain those results?  

There are terms $ \propto U^{'}_{ij}$ in equation \ref{A} that could account
for such results, as noted in \cite{Alexander}. However, such terms would be 
relevant only in systems where the pressure distribution is much larger (such as liquids and covalent glasses). In the present case $U^{'}_{ij}/R_{ij} \ll U^{''}_{ij}$, so there is only a small linear contribution 
resulting from this term.
The  $K-J$ relation without the contribution from this term is shown in Figure 6.  

Since $K$ and $J$ are related,
we can use the following relations to calculate them: 
\begin{equation}
\begin{array}{rcc}
J= \int _{fp} U'(r) g_J(r) dr +J_{bg}\\
   \\
K = \int_{fp} U''(r) g_J (r) dr +K_{bg}
\end{array}
\end{equation}
Expanding the potentials  around some  
arbitrary reference point $R_0$, the first order estimates for $J$ and $K$
 are:
\begin{equation}
\begin{array}{rcc}
J = z(J) [U'(R_0)+U''(R_0)\delta R(J)] + J_{bg}\\ 
\\
K = z(J)[U''(R_0)+U'''(R_0)\delta R(J)] + K_{bg} 
\end{array}
\end{equation}
where $\delta R(J) = R(J)_{av}-R_0$.
Using eq. 16 we can estimate the $K-J$ curve to be:
\begin{equation}
\label{K-J}
K = (J-J_{bg})F2(U(R_0))+ z(J) F1(U(R_0)) + K_{bg}
\end{equation}
with
$$F1(U(R_0))=\frac{[U''(R_0)]^2- U'(R_0)U'''(R_0)}{U''(R_0)}$$
$$F2(U(R_0)=\frac{U'''(R_0)}{U''(R_0)}$$

A natural choice for $R_0$ in eq. 17 is $R_0=R(J)_{av}$.
The results for $F1(U(R))$ and
$F2(U(R))$ are given in Figures 7a  and 7b, respectively. 
Though the corrections due to  $z(J)$, $F1(U(R(J)_{av}))$ and $F2(U(R(J)_{av}))$ are quite large, the contributions of those three large dependencies
cancel each other to create a linear curve ! 
The error of the obtained linear fit (shown in Figure 6) is less than one percent.

Thus, the form of the potential is closely related to the 
structural information given by the coordination numbers.
{\bf Moreover, the linearity of $K-J$ relation constrains the possible
 structure for a given potential}.
This holds both for glasses and for liquids (with our LJ potential). 
\begin{figure}
\bbox {(a)\epsfxsize=3.2truecm\epsfbox{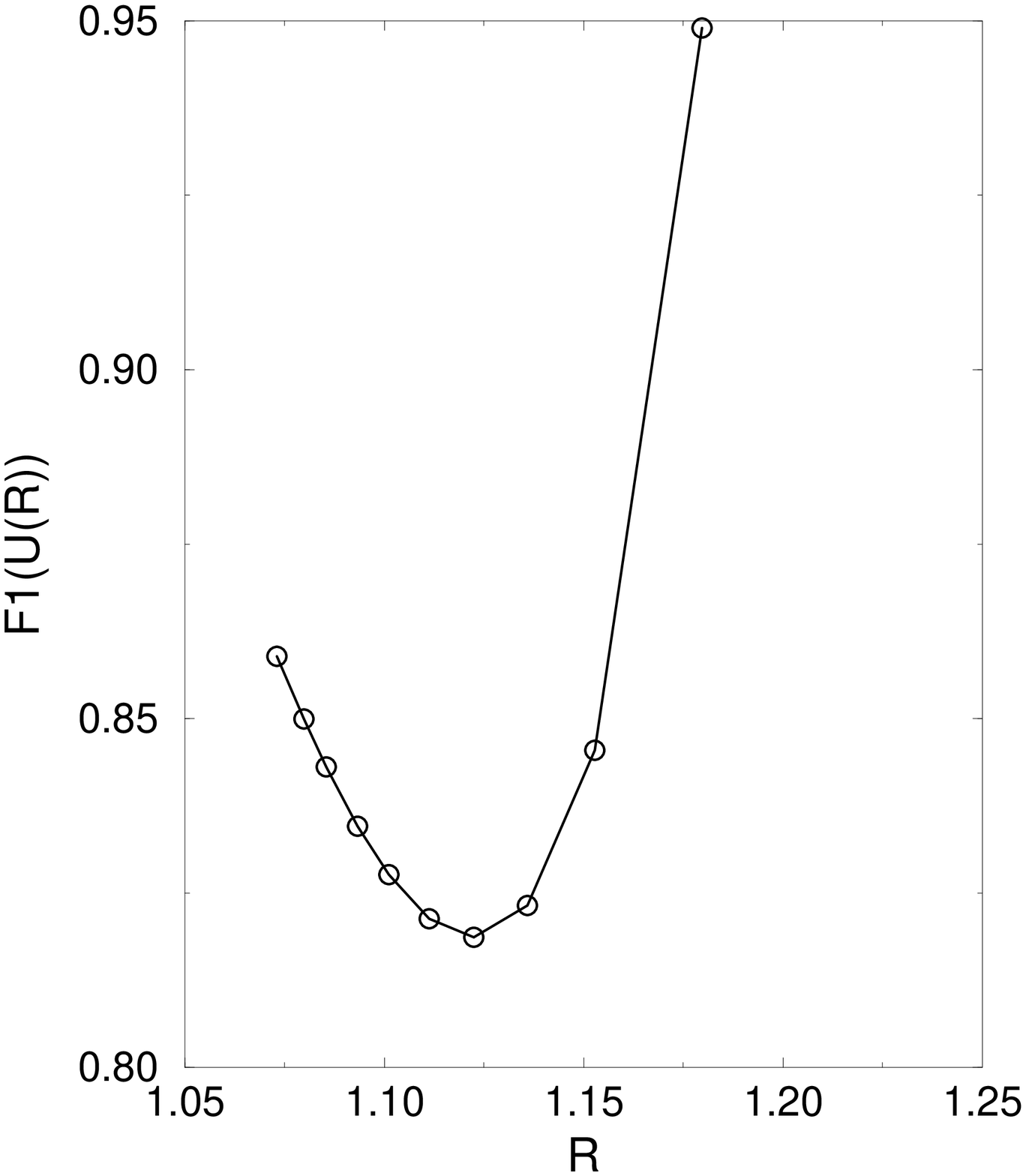}
(b)\epsfxsize=3.2truecm\epsfbox{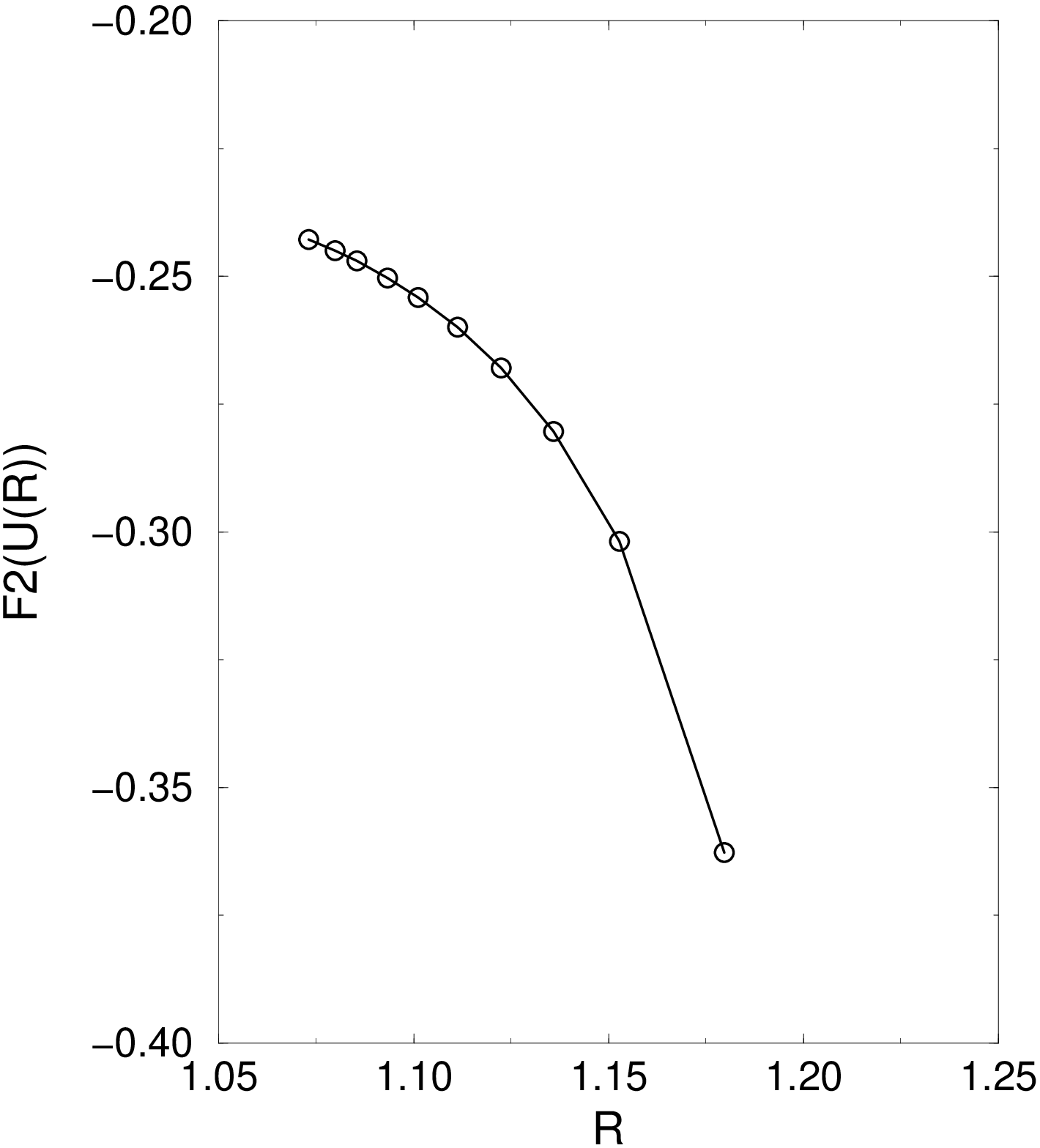}}
\vskip 0.2cm
\caption{
(a) $F1(U(R))$ vs. $R$ calculated from eq. 17 for a glass at 0K. $R$ is given in units of $\sigma$ and F1(U(R)) is in units of $\varepsilon/\sigma$.
(b)  $F2(U(R))$ vs. $R$ calculated from the same equation. }
\end{figure}

To check the extent of those results for a more general potential, we simulated a problem with different two-body potentials \cite{remark}.
In preliminary simulations we have found that the linear laws relating $K$ and $J$ (as well as $A$ and $S$) are conserved.
However, the dependence of $z(J)$ on $J$ is different, it is reduced with the coordination number.
Thus, it seems that {\bf the linear $K-J$ dependence is a more
basic law}. Obviously, this hypothesis should be studied with a set of different potentials for two-body interactions, so as to get a deeper theoretical understanding.

%
%
\section{Conclusions}

In this paper we have shown how the structural and elastic properties of 
short range mono-atomic glasses are related. 
The local pressure has a nontrivial relation to both the radius and the coordination number.
 We believe that those relations are a result of the relation between
the elastic constants and the local pressure.
Our major 
 finding is that the traces of the elastic constants have a linear relationship to the local stresses.
This holds in glasses and is even more apparent in liquids. 
We have shown that an explanation for the linear relationship lies in the 
mutual cancellation of contributions from the structural arrangement of atoms and from the potential. 
Our observations indicate that the same distorted environments exist
in the liquid and in the glass. In the liquid they are more distorted due to temperature, but still the same basic relation holds. 
It seems that the square dependence of the coordination number on the pressure is a result of this linear law. 

Our present results shed some light on the interrelation between local structural and elastic properties in simple mono-atomic glasses. In further research, a satisfactory theory for the dependence of the coordination number on $J$ and for the linearity of the $K-J$ curve should be derived using the knowledge of correlations of specific local pressures. 

This research is also related to some major unresolved issues in the physics of glasses. 
The obtained results for short range effects imply also that 
there will be longer range effects. 
The relevance of this implication for the problem of medium and long range 
correlations in glasses is discussed in a forthcoming paper \cite{Olam99}.

Knowledge of the elastic constants can provide a lot of information on various 
other aspects of glassy and liquid dynamics. In glasses, the global spectra and 
the dependence of the scattering on time/frequency is a major aspect of the 
analysis of the spectra.
The local elastic constants determine the scale of the interactions in the global dynamical matrix, which is used to solve the global phonon spectra. 
Thus, the global dynamics is in fact determined by the local disorder. 
Effects observed in the global phonon spectra, such as the boson peak and very strong scattering seen in the heat conductivity, are all manifestation of this local disorder.



\subsection*{Acknowledgments}
We thank the late S. Alexander for many illuminating discussions about the
subject. We also thank M. Flor, R. Zeitak and I. Proccacia for their comments
on the paper. 


\begin{thebibliography}{99}
\bibitem {remark-1}
One could argue for this dependence in the following manner: The area of the nearest neighbor shell
is proportional to the square of the radius  $S = R(J)_{av}^2$. If the number of atoms in the shell is proportional to the radius, we get: 
 $z(J)\sim S$. This approximation fails since the corrections due to stresses are very small as can be seen in Figure 2.
\bibitem{remark}
 We did the same calculations on a toy model of 4096 atoms interacting with a potential $U(R)=R^{-6}-A \: \exp(-R/\lambda) $ 
where $A=6\:\lambda \:\exp(1./\lambda)$ and $\lambda=0.3$.
The "atomic" mass is taken as unity and atomic spacing is taken as 0.851
(this value is found numerically to ensure that there is no pressure at equilibrium).
We used a steepest decent quench to zero temperature glass.
\end{thebibliography}
\end{document}